\documentclass[aps,prl,twocolumn,superscriptaddress,10pt,floatfix,amsmath,amssymb]{revtex4-2}
\usepackage{graphicx} 
\usepackage{dcolumn}
\usepackage{xcolor}
\usepackage{bm}
\usepackage{physics}
\usepackage{lipsum}  
\usepackage{hyperref}
\usepackage[normalem]{ulem}
\usepackage{braket}
\usepackage{orcidlink}
\usepackage{balance}
\graphicspath{{images_PRL/}}

\newcommand{\sect}[1]{\textit{#1}.--}

\newcommand{\beginsupplement}{%
        \setcounter{table}{0}
        \renewcommand{\thetable}{S\arabic{table}}%
        \setcounter{figure}{0}
        \renewcommand{\thefigure}{S\arabic{figure}}%
        \setcounter{equation}{0}
        \renewcommand{\theequation}{S\arabic{equation}}%
        \setcounter{page}{1}
        \renewcommand{\thepage}{S\arabic{page}}%
     }

\begin{document}

\title{Energetic Cost of Temporal Information Processing in Quantum Reservoirs}

\date{\today}
\author{Gabriele Cenedese}
    \email{gcenedese@ifisc.uib-csic.es}
    \affiliation{Instituto de Física Interdisciplinar y Sistemas Complejos (IFISC), UIB–CSIC UIB Campus, Palma de Mallorca, E-07122, Spain}

\author{Gonzalo Manzano}
    \email{lalo@ifisc.uib-csic.es}
    \affiliation{Instituto de Física Interdisciplinar y Sistemas Complejos (IFISC), UIB–CSIC UIB Campus, Palma de Mallorca, E-07122, Spain}

\author{Gian Luca Giorgi}
    \email{gianluca@ifisc.uib-csic.es}
    \affiliation{Instituto de Física Interdisciplinar y Sistemas Complejos (IFISC), UIB–CSIC UIB Campus, Palma de Mallorca, E-07122, Spain}

\author{Roberta Zambrini}
    \email{roberta@ifisc.uib-csic.es}
    \affiliation{Instituto de Física Interdisciplinar y Sistemas Complejos (IFISC), UIB–CSIC UIB Campus, Palma de Mallorca, E-07122, Spain}

\begin{abstract}
Quantum reservoir computing offers a promising route toward energy-efficient machine learning by processing temporal information with minimal training overhead. Yet, the physical principles linking its energetic cost to computational performance remain largely unexplored.
Here we show that, in an interacting spin reservoir, information encoding and information processing are governed by distinct physical mechanisms. In the weak interacting regime, we derive an analytical expression for the average (switching) work, showing that the energetic cost of encoding new inputs is determined by the local response of the reservoir units. In contrast, interactions primarily redistribute the encoded information, generating memory and nonlinear features while only weakly affecting the work. This separation produces opposite correlations between energetic cost and performance for representative linear and nonlinear benchmark tasks. Our results identify the switching work as the energetic signature of information encoding and clarify when energetic efficiency and computational performance are compatible.
\end{abstract}
\maketitle
\sect{Introduction} Reservoir computing (RC) has emerged as an effective approach to temporal information processing by exploiting the natural dynamics of a physical system, where training is reduced to optimizing a linear readout via linear regression while leaving the internal dynamics unchanged~\cite{Maass2002,Lukosevicius2009}. Its extension to the quantum domain has attracted increasing attention in recent years, motivated by the rich dynamics of interacting quantum systems and by the possibility of exploiting coherence, many-body quantum correlations, and dissipation for information processing~\cite{FujiiNakajima2017,Mujal2021Review}. Quantum reservoir computing (QRC) has been investigated across a broad range of physical platforms, including spin networks~\cite{FujiiNakajima2017, MartinezPena2021DPT,kornjaca2024}, fermionic lattices~\cite{Ghosh2019, kobayashi2026,llodra2023,ghosh2020reconstructing}, oscillators~\cite{Nokkala2021Gaussian,Govia2021,Kalfus2022}, photonic architectures~\cite{GarciaBeni2023,Paparelle,faccio,cimini}, superconducting circuits~\cite{Suzuki2022,chen2020,hu2024}, spin-boson platforms~\cite{Dudas2023, senanian2024,das2026}, nuclear-spin ensembles~\cite{Negoro2018}, molecules~\cite{hou2026}, and atomic arrays~\cite{AraizaBravo2022,Llodra2025Atomic}. These systems have been successfully applied to a wide range of computational classical \cite{wringe2025} and quantum tasks, including temporal learning, nonlinear information processing, forecasting of chaotic time series, and quantum-state reconstruction.

Despite this rapid development, the physical resources responsible for the computational performance of quantum reservoirs are still not fully understood, specially in what respect its energetic costs. Successful information processing requires both fading memory and nonlinear transformations, yet how these properties emerge from the underlying quantum dynamics remains an open question. Although, previous studies have related reservoir performance to dynamical phase transitions~\cite{MartinezPena2021DPT,Llodra2025Atomic,kobayashi2026}, quantum coherence~\cite{Palacios2024}, measurement protocols~\cite{Mujal2023Measurements,oriol2026}, and the role of continuous dissipation in sustaining the reservoir dynamics~\cite{Sannia2024}, how these contribute to information encoding and processing remains only partially understood. Clarifying these roles is therefore important for developing a more physical understanding of QRC and for guiding the design of more efficient reservoir architectures.

This question naturally acquires a thermodynamic perspective. In continuously driven quantum reservoirs, every new input modifies the Hamiltonian and performs work on the reservoir, which we dub ``switching work", while interaction with the environment leads to a continuous dissipation of heat. 
Since the reservoir dynamics governs both the computational response and the energetics, one may expect the two to be closely connected. Indeed, it has recently been shown that, under weak driving, predictive performance and irreversible work are simultaneously enhanced near a quantum critical region, where the closing many-body gap produces a resonance between the reservoir dynamics and the input signal~\cite{DingQiu2026}. 
That connection, however, was established in a specific dynamical regime, near a critical point and for a single class of task, leaving open whether it extends to the generic, non-critical operating regimes of QRC and whether it holds uniformly across tasks with different memory and nonlinearity requirements.

In this Letter, we show that the computational response of an interacting quantum reservoir is governed by two distinct physical mechanisms with different thermodynamic roles. The first is the direct response of the reservoir to the external input, quantified by the susceptibility of the observable coupled to the driving field. This response has a strong impact on the switching work and, when interaction effects remain perturbative, allows us to derive analytical expressions. The second originates from interactions within the reservoir. Rather than simply amplifying the response to the input, these interactions actively drive the transport of locally encoded information, endowing the system with nonlinear memory effects that become accessible to the readout while only marginally altering the work required. This distinction allows to explain the relation between energetic cost and computational performance in different tasks.

\sect{Thermodynamic framework} We consider an interacting quantum reservoir driven by a discrete input sequence. Following each new input injection, during the $k$-th input interval, the Hamiltonian is constant and can be decomposed into local and interaction terms as $\mathcal{H}_k=\sum_{i=1}^{N}\mathcal{H}_{k,i}^{\rm loc}+\mathcal{H}^{\rm int}$, where the input acts only on the local Hamiltonians $\mathcal{H}_{k,i}^{\rm loc}$, while the interaction term $\mathcal{H}^{\rm int}$ remains fixed throughout the evolution. The reservoir is weakly coupled to a thermal bath at a fixed temperature and evolves during each input interval according to the Gorini--Kossakowski--Lindblad--Sudarshan (GKLS) master equation~\cite{Gorini1976,Lindblad1976,Breuer2002} $\dot{\rho}=\mathcal{L}_k\rho=-i[\mathcal{H}_k,\rho]+\sum_{\omega}\gamma_{\omega}\mathcal{D}[L_{\omega,k}]\rho$, with dissipators $\mathcal{D}[L]\rho=L\rho L^{\dagger}-\frac{1}{2}\{L^{\dagger}L,\rho\}$. Here, the jump operators $L_{\omega,k}$ are defined as the eigenoperators of the system Hamiltonian, $[\mathcal{H}_k,L_{\omega,k}] = -\omega L_{\omega,k}$, associated with the bath-induced transitions of frequency $\omega$, while the rates $\gamma_{\omega}$ satisfy the local detailed balance condition $\gamma_{\omega} = \gamma_{-\omega} ~e^{\beta \omega}$ at the bath inverse temperature $\beta$. Their explicit construction is reported in the Supplemental Material (SM). The dissipative dynamics ensures the contractivity required for the echo-state and fading-memory properties of the reservoir~\cite{Sannia2024}. Within the standard weak-coupling framework of quantum thermodynamics, the internal energy $\mathcal{U}(t)=\Tr[\mathcal{H}(t)\rho(t)]$ satisfies the first law $\dot{\mathcal{U}}=\dot{\mathcal{W}}+\dot{\mathcal{Q}}$, where $\dot{\mathcal{W}}=\Tr[\dot{\mathcal{H}}(t)\rho(t)]$ and $\dot{\mathcal{Q}}=\Tr[\mathcal{H} (t) \dot{\rho}(t)]$ denote the work and heat currents, respectively~\cite{Alicki1979,Kosloff2013,Breuer2002,Vinjanampathy2016}. The second law of thermodynamics is manifested through the non-negativity of the entropy production (rate), $\dot{\Sigma} \equiv \dot{S} - \beta \dot{\mathcal{Q}} \geq 0$, where $\dot{S}$ is the change in the von Neumann entropy of the reservoir, which is guaranteed by the structure of the thermal GKLS generator~\cite{Davies1974,spohn1978irreversible,Spohn1978}. 

For the piecewise-constant driving protocol considered here, $\mathcal{H}_k$ remains constant over the interval $(t_{k-1},t_k)$, and changes only when a new input (instantaneously) updates the Hamiltonian from $\mathcal{H}_k$ to $\mathcal{H}_{k+1}$. Consequently, work is only performed at the times when the Hamiltonian is switched. Since the reservoir state is assumed to be unchanged during these instantaneous quenches, the corresponding {\it switching work} for input $k$ is
\begin{equation}
    \mathcal{W}_k
    =
    \Tr\!\left[
        \left(
            \mathcal{H}_{k+1}-\mathcal{H}_k
        \right)
        \rho_k
    \right],
    \label{eq:general_switching_work}
\end{equation}
where $\rho_k=\rho(t_k)$ denotes the reservoir state immediately before the update. Strictly speaking, $\rho_k$ depends on the entire history of previous inputs, but this dependence is omitted for notational simplicity. This switching work represents the energetic cost of injecting a new input into the reservoir and will be the central quantity of our analysis. 
On the other side, the reservoir is continuously exchanging energy with the environment in the form of heat, thus leading to dissipation and changes in the reservoir entropy. The heat associated to the interval $(t_{k-1},t_k)$, during which $\mathcal{H}_k$ remains constant, reads: 
\begin{equation}
    \mathcal{Q}_k
    =
    \int_{t_{k-1}}^{t_k}dt\,
    \Tr\!\left[
        \mathcal{H}_k \sum_{\omega}\gamma_{\omega}\mathcal{D}[L_{\omega,k}]\rho(t)
    \right].
    \label{eq:heat_interval}
\end{equation}
Combining the continuous evolution with the instantaneous update, the first law over one computational step reads $\Delta\mathcal{U}_k=\mathcal{Q}_k+\mathcal{W}_k$, and the second law becomes $\Sigma_{k} = \beta (\mathcal{W}_k - \Delta \mathcal{F}_k) \geq 0$, which imposes that the switching work needed for injecting a new input must overcome the change in (non-equilibrium) free energy~\cite{parrondo2015thermodynamics} of the reservoir $\Delta \mathcal{F}_k \equiv \Delta\mathcal{U}_k - \Delta S_k/\beta$, during the computational step. 
After an initial washout, during which the reservoir loses memory of its arbitrary initial condition, energetic quantities can be averaged over different input steps. Since we assume that the inputs are independently sampled from a stationary probability distribution, the reservoir reaches a statistically stationary regime, implying $\langle\Delta\mathcal{U}_k\rangle=0$, and $\langle\Delta S_k\rangle=0$, and therefore $\langle\mathcal{W}_k\rangle=-\langle\mathcal{Q}_k\rangle$ and $\langle \Sigma_k \rangle = \beta \langle \mathcal{W}_k\rangle$, where $\langle\cdot\rangle$ denotes the average taken over the input history. The average energetic balance and dissipation can thus be fully characterized through the switching work, which will be the focus of the remainder of this work.

Equation~\eqref{eq:general_switching_work} identifies the energetic cost associated with updating the reservoir with a new input. Its value depends on the reservoir state established during the preceding input interval. In the following, we show that, in the perturbative interaction regime, the average switching work is fully determined by the stationary response of the uncoupled reservoir units, even though interactions may substantially enhance memory and nonlinear processing by redistributing the information already encoded in the reservoir.

\sect{Model and analytical results} We consider a reservoir composed of $N$ interacting spin-$1/2$ particles driven by a discrete input sequence continuously coupled to a thermal bath, with dynamics governed by a global GKLS master equation. During the $k$-th input interval of duration $\Delta t$, the Hamiltonian is
\begin{equation}
    \mathcal{H}_k
    =
    \sum_{i=1}^{N}
    \left[
        h_i \sigma_i^z
        +
        a_k \sigma_i^x
    \right]
    +
    \sum_{i<j}
    J_{ij}\sigma_i^x\sigma_j^x ,
    \label{eq:model_hamiltonian}
\end{equation} 
where the input is encoded through the local field $a_k=A(s_k+1)$. Here, $A$ sets the overall amplitude of the input modulation, while the dimensionless variable $s_k$ is sampled from the input distribution. The transverse local fields $h_i=h+\delta_i$, contain static detunings sampled from a uniform distribution, $\delta_i\sim\mathcal{U}[-D,D]$ which break the homogeneity of the reservoir. The interaction strengths are also independently sampled from $J_{ij}\sim\mathcal{U}[-J_s,J_s]$, so that $D$ controls the local disorder and $J_s$ sets the characteristic interaction scale.
Throughout this Letter we set $A=1$ and consider the homogeneous case $D=0$, in order to isolate the dependence on the local field and interaction strength. The effects of varying the input amplitude and disorder strength are discussed in the End Matter.

The reservoir is measured at the end of each input interval to construct the readout features. Following the standard QRC protocol, the feature vector consists of the single-spin expectation values $\langle\sigma_i^\alpha\rangle$ and the two-spin correlation functions $\langle\sigma_i^\alpha\sigma_j^\alpha\rangle$ ($\alpha=x,y,z$, $i<j$), together with a constant bias.

To gain analytical insight into the switching work, we first consider the regime of weak interactions $J_s \ll \Omega_{i,k}\equiv\sqrt{h_i^2+ a_k^2}$, where the local dissipative description provides a perturbatively accurate approximation of the reservoir dynamics~\cite{Trushechkin2016,Gonzalez2017}. Throughout this work we set $\gamma\Delta t=1$, so that the duration of each input interval $\Delta t$ is comparable to the local relaxation time. Under these conditions the reservoir does not fully relax to its local stationary state after every input update, and the analytical expressions derived below should therefore be interpreted as leading-order approximations rather than exact stationary results. In particular, for vanishing interactions ($J_s=0$), small quantitative deviations are expected because the local stationary state is not completely reached within a single input interval. Interesting, introducing weak interactions is found to improve the agreement with the local analytical predictions by promoting mixing of the locally encoded information, while remaining within the perturbative regime where the local GKLS description is expected to be valid~\cite{Trushechkin2016,Gonzalez2017}.

The local Hamiltonian of the $i$-th spin is $\mathcal{H}_{k,i}^{\rm loc}=h_i\sigma_i^z+a_k\sigma_i^x$, with local level splitting $\Omega_{i,k}$. For each   single spin, the local GKLS dynamics at fixed input relaxes to the (thermal) stationary polarizations
\begin{align}
    \langle\sigma_i^x\rangle_{\rm ss}
    &=
    -
    \frac{a_k}{\Omega_{i,k}}
    \tanh(\beta\Omega_{i,k}),
    \label{eq:sx_stationary}
    \\
    \langle\sigma_i^y\rangle_{\rm ss}
    &=
    0,
    \label{eq:sy_stationary}
    \\
    \langle\sigma_i^z\rangle_{\rm ss}
    &=
    -
    \frac{h_i}{\Omega_{i,k}}
    \tanh(\beta\Omega_{i,k}).
    \label{eq:sz_stationary}
\end{align}
In the low-temperature limit ($\beta \,\Omega_{i,k}\gg1$), the stationary polarization is determined by the nonlinear local response $F(a,h)={a}/{\sqrt{h^2+a^2}}$, such that $\langle\sigma_i^x\rangle_{\rm ss}=-F(a_k,h_i)$. The function $F(a,h)$ quantifies the susceptibility of an individual reservoir unit to the input field. It is approximately linear when the input amplitude is small compared with the local field ($a_k\ll h_i$) and becomes strongly nonlinear as the two scales become comparable. Since the input modifies only the field in the $x$ direction, the work performed by the instantaneous update $a_k\rightarrow a_{k+1}$ becomes
\begin{eqnarray}
    \mathcal{W}_k
    &=&
    (a_{k+1}-a_k)
    \sum_{i=1}^{N}
    \Tr[\sigma_i^x\rho_k] \nonumber\\
    &\simeq&
    -(a_{k+1}-a_k)
    \sum_{i=1}^{N}
    F(a_k,h_i),
    \label{eq:perturbative_switching_work}
\end{eqnarray}
where $\rho_k$ is the reservoir state immediately before the switch. For statistically independent consecutive inputs uniformly distributed over $a_k\in[A,2A]$ and static disorder uniformly distributed over $\delta_i\in[-{D,D}]$, averaging Eq.~\eqref{eq:perturbative_switching_work} over both the input distribution and the disorder gives the mean work per switch
\begin{eqnarray}
    \langle{\mathcal{W}_k}\rangle\equiv\mathbb{E}_{\delta,a}[\mathcal{W}_k]
    \simeq
    \frac{N}{2DA}
    \int_A^{2A} da\,
    \left(
        a^2-\frac{3A}{2}a
    \right)\times \nonumber\\
    \times\left[
        \text{arsinh}\!\left(\frac{h+D}{a}\right)
        -
        \text{arsinh}\!\left(\frac{h-D}{a}\right)
    \right].
    \label{eq:mean_switching_work}
\end{eqnarray}
Equation~\eqref{eq:mean_switching_work} provides the analytical prediction used below to characterize the energetic cost of input injection (see the SM for the full derivation), and it assumes a continuous input distribution, for the numerical simulations below, we instead use discretized inputs and replace the integral with a finite sum. It shows that, in the non-interacting regime, the average switching work is governed by the local input response of the reservoir units. On the other hand, the interactions can still substantially modify the computational dynamics by redistributing the encoded information and generating memory and nonlinear features, although they have no direct impact on the work.
Beyond the perturbative regime, interaction-induced corrections become significant and the local description is no longer expected to provide a quantitatively accurate prediction.

\begin{figure}[h]
	\includegraphics[width=\columnwidth]{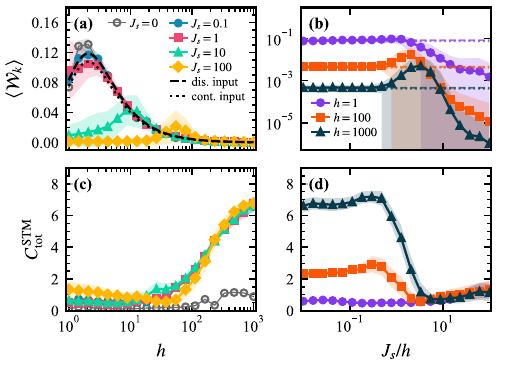}
	\caption{Average work and STM performance. (a,b) $\langle\mathcal{W}_k\rangle$ as a function of $h$ for different $J_s$ and of $J_s/h$ for different $h$, respectively, numerical results are additionally averaged over independent couplings realizations. (c,d) Total STM capacity, $C_{\rm tot}^{\rm STM}$, as a function of $h$ for different $J_s$ and of $J_s/h$ for different $h$, respectively. Dashed lines in panels (a) and (b) show the analytical prediction of $\langle\mathcal{W}_k\rangle$ for discretized inputs. The dotted line in panel (a) corresponds to the continuous-input limit as in Eq.~\ref{eq:mean_switching_work}. Shaded regions indicate one standard deviation over $100$ independent realizations of the random couplings and input sequences. Here and in the following we set $k_B = 1$ and $\hbar=1$ and we fix $\beta = 10$ (in inverse energy units).}
	\label{fig:1}
\end{figure} 
\sect{Numerical testing} We now compare the analytical predictions with the computational performance of the quantum reservoir. We consider three benchmark tasks probing complementary computational capabilities. For each task, the performance is quantified by the corresponding capacity,
$
C=\mathrm{Cov}(\bm{y},\bm{\hat y})^2/\left[\sigma^2(\bm{y})\sigma^2(\bm{\hat y})\right],
$
where $\bm{y}$ denotes the target sequence and $\bm{\hat y}$ the optimized linear readout. The first task corresponds to a short-term memory (STM) task consisting in the reconstruction of a delayed input, with target $y_k=s_{k-\tau}$, with $\tau$ the delay. Its performance is quantified by the total memory capacity $C_{\rm tot}^{\rm STM}=\sum_\tau C_\tau^{\rm STM}$ \cite{Jaeger2002,Lukosevicius2009, wringe2025}. The second task is the parity-check (PC) which probes nonlinear information processing through the target $y_k=\left(\sum_{m=0}^{n-1}s_{k-m}\right)\bmod 2$ \cite{FujiiNakajima2017}; throughout the main text we consider $n=4$ and report the corresponding capacity $C^{\rm PC}_4$. Finally, we consider the nonlinear autoregressive moving-average (NARMA) task, which requires the reservoir to reproduce the dynamics $y_{k+1}=0.3y_k+0.05y_k\sum_{m=0}^{n-1}y_{k-m}+1.5\,s_{k-n+1}s_k+0.1$ \cite{Atiya2000,Lukosevicius2009}. We show the fifth-order benchmark and report the corresponding capacity $C^{\rm NARMA}_5$. Additional PC and NARMA orders are presented in the End Matter.

Unless otherwise specified, all results are obtained for a reservoir of $N=5$ spins. Each realization consists of $1000$ washout, $1000$ training, and $1000$ testing steps, with a linear readout trained on the observables introduced above. All reported quantities are averaged over $100$ independent realizations of both the random couplings and the input sequence. For the extensive parameter scans, the STM and NARMA inputs are discretized into $21$ equally spaced values, whereas the PC task naturally employs binary inputs. For the NARMA benchmark, the input sequence takes values $s_k\in [0,0.2]$ to avoid instabilities of the target dynamics.
\begin{figure}[t!]
\includegraphics[width=\columnwidth]{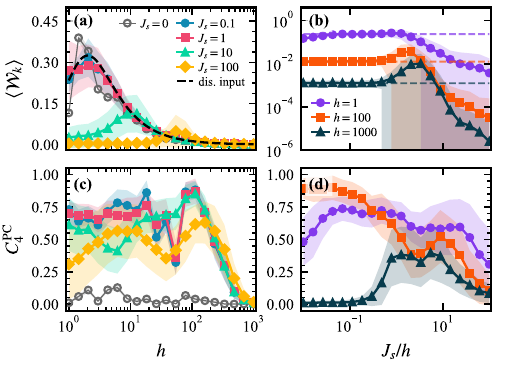}
	\caption{Same as Fig.~\ref{fig:1} for the fourth-order PC capacity, $C^{\rm PC}_4$.} 
	\label{fig:2}
\end{figure} 

To investigate the interplay between energetics and computation, we perform two complementary parameter scans. In the first, the local field is varied over three orders of magnitude, $h\in[1,10^3]$ (here and throughout, all energy scales $(h, J_s, A, D)$ are expressed in the same arbitrary energy units), for different interaction strengths $J_s\in\{0,10^{-1},10^0,10^1,10^2\}$. In the second, the local field is fixed ($h\in\{1,10^2,10^3\}$) while the relative interaction strength is varied over the range $J_s/h\in[10^{-2},10^2]$. This enables us to separately investigate the roles of the local input response and interaction-induced information redistribution in both the weak and strong intra-coupling regimes.
\begin{figure}[t!]	
\includegraphics[width=\columnwidth]{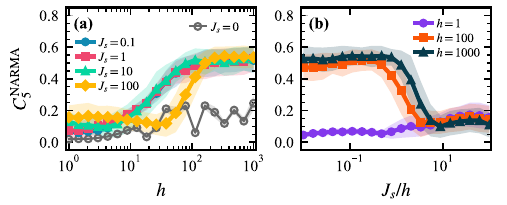}
	\caption{Same as Fig.~\ref{fig:1}(c,d) for the fifth-order NARMA capacity, $C^{\rm NARMA}_5$. The switching work follows Fig.~\ref{fig:1}(a,b), differing only by the input rescaling used for the NARMA benchmark.}
	\label{fig:3}
\end{figure} 

Fig.~\ref{fig:1}(a) and \ref{fig:2}(a) validate the analytical expression for the average switching work, Eq.~\eqref{eq:mean_switching_work}. The analytical prediction accurately reproduces the numerical results in the perturbative regime ($J_s/h\ll1$), becoming essentially exact for $J_s=10^{-1}$ and remaining an excellent asymptotic approximation up to $J_s/h\sim1$. The STM capacity increases monotonically with the local field [Fig.~\ref{fig:1}(c)] for $h \gtrsim10^2$. This behavior originates from the progressive linearization of the local response: for $h\gg a_k$, one has $F(a_k,h)\simeq a_k/h$, while nonlinear corrections are suppressed as $O(a_k^3/h^3)$ (see SM). The input is therefore encoded more linearly, enhancing the reservoir's linear memory while maintaining a low dissipation and work cost. The same trend is observed for the NARMA benchmark [Fig.~\ref{fig:3}(a)], also requiring significant memory while placing relatively weak demands on nonlinear processing. 

By contrast, the PC task displays the opposite behavior [Fig.~\ref{fig:2}(c)]. Its capacity increases as $h$ decreases and the local response becomes increasingly nonlinear. This trend persists down to $h\simeq100$, below which the behavior changes qualitatively. In this regime, the capacity develops oscillations as a function of $h$, whose characteristic period depends on the interaction strength, with clear aliasing effects for weak interactions. At the same time, the capacity decreases with increasing $J_s$ as can be appreciated even if the dynamics is oscillatory [Fig.~\ref{fig:2}(c)]. This suggests a crossover from a regime dominated by local nonlinear encoding to one in which interactions increasingly redistribute the encoded information. While this redistribution enriches the reservoir dynamics, it progressively reduces the contribution of the local nonlinear encoding that is most beneficial for this task.

The complementary scans as a function of $J_s/h$ further support this picture. As shown in Figs.~\ref{fig:1}(b) and \ref{fig:2}(b), the average switching work remains essentially independent of the interaction strength for $J_s/h\lesssim1$, confirming the analytical prediction. Around $J_s/h\sim1$, however, the work exhibits a small maximum before rapidly decreasing to zero as the interaction-dominated regime is approached. The computational performance distinguishes these two dynamical regimes. For STM [Fig.~\ref{fig:1}(d)] and NARMA [Fig.~\ref{fig:3}(b)], the capacity remains approximately constant throughout the weak intra-coupling regime and then sharply decreases once $J_s/h\gtrsim1$, indicating the loss of useful memory. The plateau values are anti-correlated with the switching work, showing that the energetic cost alone does not determine the computational performance. The PC benchmark displays a richer behavior (Fig.~\ref{fig:2}(d)): for $h=1$ and $h=10^2$ the capacity exhibits an overall decrease with increasing interaction strength. By contrast, for $h=10^3$ it develops a small maximum around the crossover $J_s/h\sim1$, coinciding with the bump in the switching work. Interestingly, a qualitatively similar enhancement near the crossover was also reported in \cite{DingQiu2026}.
\begin{figure}[t!]	
\includegraphics[width=\columnwidth]{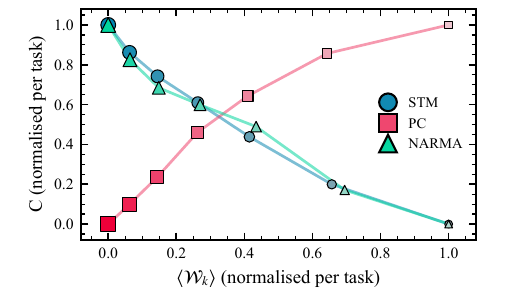}
	\caption{Normalized task capacity versus normalized average switching work per input for the STM, fourth-order PC, and fifth-order NARMA tasks, obtained by varying $h$ from $10^2$ to $10^3$. For each task, both the capacity and the average switching work are independently rescaled to the interval $[0,1]$. Circles, squares, and triangles denote the STM, PC, and NARMA tasks, respectively; marker size and color saturation increase with $h$.} 
	\label{fig:4}
\end{figure} 
Additional results reported in the End Matter further support our interpretation of the interplay between information encoding, energetics, and computational performance. In particular, the same relationship between the linearity of the local encoding and task performance is observed upon varying the disorder strength, whereas increasing the input amplitude mainly increases the energetic cost while only weakly affecting the computational performance.

\sect{Conclusions} In this work we address the physical principles relating QRC energetic cost to computational performance. We establish numerically and analytically the connection between the energetic cost of information injection and the local encoding mechanism in QRC. This relation accurately predicts the switching work over the perturbative interaction regime and provides a physical framework to interpret the computational behavior of different tasks. Fig.~\ref{fig:4} summarizes the main result of this work: the energetic-performance trade-off in QRC is not universal, but depends on the type of information processing required by the task. Rather than displaying a universal relation with computational performance, the switching work quantifies the energetic cost of encoding information into the reservoir, while the computational advantage is determined by whether the task requires a linear or nonlinear memory of the input. In this setting, we find that more linear tasks, such as STM and NARMA, benefit from an increasingly linear response at a marginal work cost, whereas the PC task achieves its best performance in the regime where the local response is strongly nonlinear, which is also associated with higher energetic costs. Consequently, Fig.~\ref{fig:4} reveals opposite energetic-performance trade-offs: STM and NARMA exhibit a negative correlation between capacity and switching work, while PC displays a positive one. This shows that the energy-performance connection reported near criticality in \cite{DingQiu2026} is not a generic feature of QRC, but reflects the specific linear/nonlinear requirements of the task at hand.

An interesting direction for future work is to extend this analysis for increasingly complex tasks on the same inputs and going beyond specific benchmark tasks by combining the present thermodynamic framework with the information processing capacity \cite{Dambre2012, MartinezPena2023IPC}. This would provide a broader characterization of how energetic resources are distributed between linear and nonlinear information processing. Furthermore, addressing the  energetics of different protocols would allow to establish more general relations with the performance of QRC. More broadly, our framework provides a route towards energetically informed design principles for quantum reservoirs, where the encoding mechanism and the computational task can be jointly optimized. Further costs involved in the reservoir measurement and final information readout may also be considered~\cite{Sagawa09,Guryanova20,Latune25,VanVu25}, as well as comparison with the thermodynamic costs of computation in other platforms~\cite{Freitas21,Patryk24,Manzano24}. Spanning these ideas beyond the perturbative regime and to experimentally realistic dissipative architectures represents a natural direction for future research.

\begin{acknowledgments}
\sect{Acknowledgments} We thank Oriol Morguí, Nathan Keenan, José Antonio Almanza,  Luisa Toledo, and Albert Cabot for insightful discussions.
We acknowledge funding from the Spanish State Research Agency through the COQUSY project PID2022-140506NB-C21 and -C22 and the María de Maeztu project CEX2021-001164-M, and the QuantERA QNet project PCI2024-153410 and CoQuaDis project PCI2024-153446, funded by MICIU/AEI/10.13039/501100011033 and by ERDF, EU.
\end{acknowledgments}
\bibliography{biblio.bib}
\section{End matter}
\begin{figure}[h!]
    \includegraphics[width=\columnwidth]{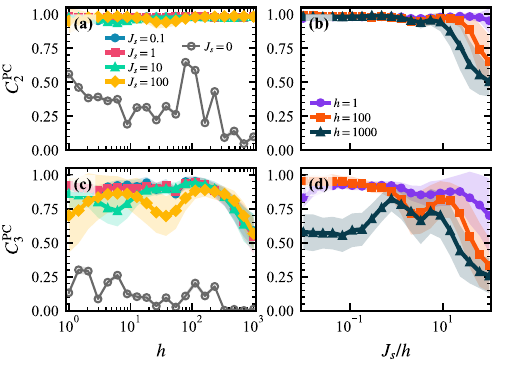}
	\caption{Average capacity for the second- and third-order PC tasks. Panels (a,c) show the dependence on the local field $h$, while panels (b,d) report the corresponding scans as a function of $J_s/h$.}
    \label{fig:5}
\end{figure}

\begin{figure}[t!]
    \includegraphics[width=\columnwidth]{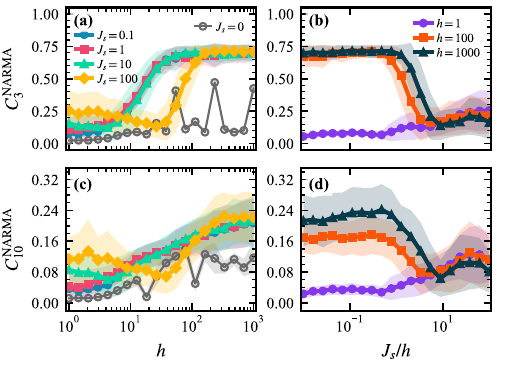}
	\caption{Average capacity for the third- and tenth-order NARMA tasks. Panels (a,c) show the dependence on the local field $h$, while panels (b,d) report the corresponding scans as a function of $J_s/h$.}
    \label{fig:6}
\end{figure}

\sect{Additional PC and NARMA benchmark} To demonstrate that the observed energetic-performance trade-off is not specific to the benchmarks discussed in the main text, we report additional results for different order PC and NARMA tasks.
Figure~\ref{fig:5} reports the second- and third-order PC capacities. The third-order task reproduces the same qualitative behavior discussed in the main text for the fourth-order PC benchmark, confirming that nonlinear tasks benefit from the nonlinear local encoding at small $h$. By contrast, the second-order PC task is sufficiently simple to achieve nearly maximal performance over most of the parameter space. The capacity is reduced only for vanishing interactions, where memory is suppressed, and for sufficiently strong interactions, where the interaction-dominated regime suppresses the local encoding mechanism.
Figure~\ref{fig:6} reports the third- and tenth-order NARMA capacities. Both tasks display the same qualitative trends observed for the fifth-order benchmark discussed in the main text, confirming that linear-memory tasks benefit from the asymptotically linear local encoding. As expected, the tenth-order task exhibits an overall degradation of the computational performance due to its increased complexity.

\begin{figure}[t!]
    \includegraphics[width=\columnwidth]{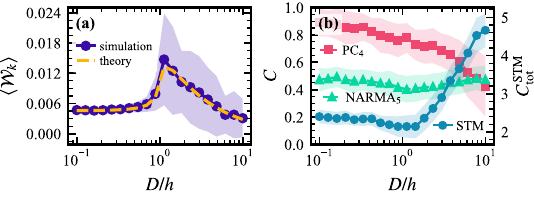}
	\caption{(a) Average switching work as a function of the disorder strength $D/h$. The dashed line shows the analytical prediction of Eq.~\eqref{eq:mean_switching_work}. (b) Average STM, fourth-order PC, and fifth-order NARMA capacities as a function of $D/h$.}
    \label{fig:7}
\end{figure}

\sect{Scan over disorder strength} Figure~\ref{fig:7} reports the dependence of the computational performance and switching work on the disorder strength $D$, for fixed $J_s/h=10^{-2}$ and $A/h=10^{-2}$. For weak disorder ($D/h\ll 1$), the reservoir behavior remains essentially unchanged. This is consistent with the analytical expansion of the disorder-averaged response, where the leading correction is quadratic in $D$, namely $\mathbb{E}_\delta[F(a,h+\delta)]=F(a,h)+O(D^2/h^3)$ (see SM). As the disorder strength increases, the local response progressively enters the strong-disorder regime, where
\begin{equation}
    \mathbb{E}_\delta[F(a,h+\delta)]\simeq \frac{a}{D}\ln\!\left(\frac{2D}{a}\right)\sim\frac{\ln D}{D}a,
\end{equation}
up to subleading corrections (see SM). The response therefore becomes asymptotically linear in the input amplitude. Consequently, the same trends discussed in the main text are recovered: the STM capacity increases, whereas the PC capacity decreases as the nonlinear local encoding is progressively suppressed. Finally, the analytical prediction for the switching work remains in excellent agreement with the numerical results throughout the explored parameter range.

\begin{figure}[t!]
    \includegraphics[width=\columnwidth]{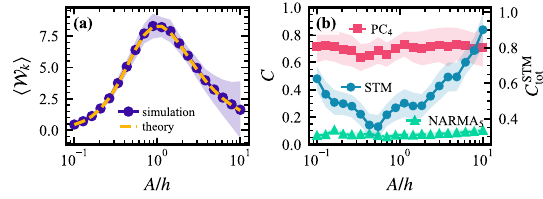}
	\caption{(a) Average switching work as a function of the input amplitude $A/h$. The dashed line shows the analytical prediction of Eq.~\eqref{eq:mean_switching_work}. (b) Average STM, fourth-order PC, and fifth-order NARMA capacities as a function of $A/h$.}
    \label{fig:8}
\end{figure}

\sect{Scan over input strength} Figure~\ref{fig:8} shows the dependence of the computational performance and switching work on the input amplitude $A/h$, for fixed $J_s/h=10^{-2}$ and $D=0$. Over the explored parameter range, varying the input strength has only a minor effect on the computational performance. By contrast, the switching work changes by several orders of magnitude, increasing for $A/h\ll1$, reaching a maximum around $A/h\sim1$, and subsequently decreasing in the strong-driving regime. This behavior follows from the local response $F(a,h)$ which is linear for $a/h\ll1$ and saturates as $F(a,h)\simeq1-h^2/(2a^2)$ for $a/h\gg1$ (see SM). The marked variation of the energetic cost in the absence of a comparable change in computational performance provides further evidence that the switching work is not, in general, a direct predictor of reservoir performance. Additional scans for $D$ and $A$ for different values of $J_s/h$ are reported in the SM.
\newpage
\clearpage
\onecolumngrid
\beginsupplement
\vspace*{1em}
\begin{center}
{\large\bfseries Supplemental Material for\par}
\vspace{0.5em}
{\large\bfseries ``Energetic Cost of Temporal Information Processing in Quantum Reservoirs''\par}
\end{center}
\vspace{2em}

This Supplemental Material provides the technical details underlying the results presented in the main text. We describe the microscopic construction of the global GKLS master equation, derive the analytical expression for the switching work associated with the piecewise-constant driving protocol, and report additional numerical results.

\section*{Microscopic description of the thermal GKLS generator}
\label{app:global_master_equation}

In the main text, the reservoir dynamics is described by a thermal GKLS \cite{Gorini1976,Lindblad1976} master equation whose jump operators are associated with the instantaneous interacting Hamiltonian. In this section we present its microscopic construction within the standard weak-coupling theory of open quantum systems \cite{Breuer2002}.

The reservoir is driven by a piecewise-constant input sequence. During each input interval $[t_{k-1},t_k]$, the Hamiltonian $\mathcal{H}_k$ remains fixed, so that the dynamics is governed by a time-independent open quantum system. We therefore consider the microscopic Hamiltonian
\begin{equation}
    \mathcal{H}_{\rm tot}
    =
    \mathcal{H}_k
    +
    \mathcal{H}_{\rm B}
    +
    \mathcal{H}_{\rm SB},
\end{equation}
where $\mathcal{H}_{\rm B}$ is the Hamiltonian of an equilibrium thermal bath, assumed to be initially in the Gibbs state
\begin{equation}
    \rho_{\rm B}
    =
    \frac{e^{-\beta\mathcal{H}_{\rm B}}}
    {\Tr\!\left[e^{-\beta\mathcal{H}_{\rm B}}\right]},
\end{equation}
with inverse temperature $\beta$, and
\begin{equation}
    \mathcal{H}_{\rm SB}
    =
    S\otimes B
\end{equation}
describes the system--bath interaction. Here $S$ acts on the reservoir Hilbert space, while $B$ acts on the bath degrees of freedom. Throughout this work we choose the system coupling operator
\begin{equation}
    S
    =
    \sum_{i=1}^{N}
    \sigma_i^y.
\end{equation}
This choice reflects the physical assumption that the coupling between the
reservoir and the environment is fixed by the device architecture. Accordingly, the external input modifies only the system Hamiltonian $\mathcal{H}_k$, while the system--bath interaction remains unchanged. Moreover, since $S$ does not commute with the local driving Hamiltonian $\sum_ih_i\sigma_i^z+\sum_ia_k\sigma_i^x$ for any nontrivial choice of the local parameters, the system--bath interaction always contains finite-frequency components. The bath can therefore induce energy-exchange processes throughout the entire parameter regime considered in this work, rather than reducing to purely dephasing dynamics.

The Davies construction starts from the spectral decomposition of the
interacting Hamiltonian,
\begin{equation}
    \mathcal{H}_k
    =
    \sum_{\epsilon}
    \epsilon\,
    \Pi_{\epsilon}^{(k)},
\end{equation}
which defines the Bohr frequencies $\omega=\epsilon'-\epsilon$. The system coupling operator is then projected onto the corresponding Bohr-frequency sectors,
\begin{equation}
    A_k(\omega)
    =
    \sum_{\epsilon'-\epsilon=\omega}
    \Pi_{\epsilon}^{(k)}
    S\,
    \Pi_{\epsilon'}^{(k)},
    \label{eq:jump_operator_definition}
\end{equation}
where the sum runs over all pairs of eigenstates separated by the same energy difference $\omega$. Consequently, each operator $A_k(\omega)$ collects all transitions associated with the corresponding Bohr frequency and constitutes an elementary dissipative channel entering the GKLS generator.

The reduced system dynamics is obtained under the standard assumptions of weak system--bath coupling, an initially factorized state $\rho_{\rm tot}(0)=\rho(0)\otimes\rho_{\rm B}$, the Born--Markov approximation, and the full secular approximation. The weak coupling assumption ensures that the bath only weakly perturbs the system dynamics, while the Born approximation neglects system--bath correlations generated during the evolution. The Markov approximation assumes that the bath correlation time is much shorter than the characteristic relaxation time of the system, so that memory effects induced by the environment can be neglected. Finally, the secular approximation discards rapidly oscillating terms coupling different Bohr frequencies, yielding a completely positive dynamical semigroup \cite{Gorini1976,Lindblad1976,Breuer2002,Davies1974}.
Under these assumptions, the reduced dynamics is governed by the thermal GKLS generator
\begin{align}
    \mathcal{L}_k\rho
    ={}&
    -i[\mathcal{H}_k,\rho]
    +
    \sum_{\omega}
    \Gamma(\omega)
    \mathcal{D}[A_k(\omega)]\rho,
    \label{eq:davies_general}
\end{align}
where we neglect the lamb shift term in the Hamiltonian evolution and
\begin{equation}
    \mathcal{D}[L]\rho
    =
    L\rho L^\dagger
    -
    \frac12
    \left\{
        L^\dagger L,
        \rho
    \right\}
\end{equation}
is the Lindblad dissipator. In the notation adopted in the main text, the jump operators and transition rates are simply identified as
\begin{equation}
    L_{\omega,k}\equiv A_k(\omega),
    \qquad
    \gamma_{\omega}\equiv\Gamma(\omega).
\end{equation}
The transition rates are microscopically determined by the Fourier transform of the bath correlation function,
\begin{equation}
    \Gamma(\omega)
    =
    \int_{-\infty}^{+\infty}
    dt\,
    e^{i\omega t}
    \langle
        B(t)B(0)
    \rangle,
\end{equation}
where $B(t)=e^{i\mathcal{H}_{\rm B}t}Be^{-i\mathcal{H}_{\rm B}t}$ is the bath coupling operator in the Heisenberg picture. The bath correlation function therefore encodes the spectral properties of the thermal environment. Thermal equilibrium of the bath implies the Kubo--Martin--Schwinger (KMS) relation
\begin{equation}
    \Gamma(-\omega)
    =
    e^{-\beta\omega}
    \Gamma(\omega),
    \label{eq:kms}
\end{equation}
which guarantees the local detailed balance condition adopted in the main text and ensures that the Gibbs state $e^{-\beta \mathcal{H}_k}/Z_k$ associated with $\mathcal{H}_k$ is the stationary state of the dynamics.

Throughout this work we adopt the simplest wide-band approximation by assuming a flat bath spectral density, $J(\omega)=\gamma$, independent of frequency, , yielding $\gamma_{|\omega|}=\gamma$ and $\gamma_{-|\omega|}=\gamma ~ e^{-\beta|\omega|}$.  Consequently, all downward transitions occur with the same bare relaxation rate, while the corresponding excitation rates are fixed by the KMS relation. Different choices of the bath spectral density would modify the relative relaxation rates associated with different Bohr frequencies and therefore the transient approach to the stationary state. Since all thermodynamic quantities analyzed in this work are evaluated after the an initial washout period, when the reservoir has reached its stationary operating regime, such modifications are not expected to produce qualitative changes to our results.

Finally, in the numerical implementation the jump operators are constructed according to Eq.~(\ref{eq:jump_operator_definition}) for all Bohr frequencies, including the zero-frequency sector. Therefore, transitions within degenerate eigenspaces are fully retained in the dissipative dynamics.
The resulting thermal GKLS generator constitutes the microscopic model used throughout the main text to describe the dissipative dynamics of the reservoir under piecewise-constant driving

\section*{Derivation of the mean switching work}
\label{app:work_derivation}
The microscopic construction presented above provides the exact thermal GKLS generator used throughout the numerical simulations. The analytical derivation below is instead performed in the perturbative regime discussed in the main text, where the microscopic global GKLS dynamics is accurately approximated by a local description. This allows us to derive the leading-order analytical expression for the switching work.

For a single spin ($J_{ij}=0$), the instantaneous local Hamiltonian in the presence of disorder and input driving is
\begin{equation}
\mathcal{H}_{i,k}^{\mathrm{loc}}
=
h_i\sigma_i^z+a_k\sigma_i^x,
\qquad
h_i=h+\delta_i,
\qquad
a_k=A(s_k+1).
\end{equation}
It can be diagonalized by introducing $\Omega_{i,k}=\sqrt{h_i^2+a_k^2}$, $\omega_{i,k}=2\Omega_{i,k}$, and the angle $\cos\theta_{i,k}=h_i/\Omega_{i,k}$, $\sin\theta_{i,k}=a_k/\Omega_{i,k}$. The rotated Pauli $z$ operator along the instantaneous local field is
\begin{equation}
\tilde{\sigma}_{i,k}^{z}
=
\cos\theta_{i,k}\sigma_i^z
+
\sin\theta_{i,k}\sigma_i^x ,
\end{equation}
so that
\begin{equation}
\mathcal{H}_{i,k}^{\mathrm{loc}}
=
\Omega_{i,k}\tilde{\sigma}_{i,k}^{z}.
\end{equation}
The corresponding instantaneous eigenstates are denoted by $\ket{\tilde{0}_{i,k}}$ and $\ket{\tilde{1}_{i,k}}$, where $\ket{\tilde{0}_{i,k}}$ is the local ground state and $\ket{\tilde{1}_{i,k}}$ is the local excited state. Explicitly,
\begin{equation}
\ket{\tilde{1}_{i,k}}
=
\cos\frac{\theta_{i,k}}{2}\ket{0_i}
+
\sin\frac{\theta_{i,k}}{2}\ket{1_i},
\quad
\ket{\tilde{0}_{i,k}}
=
-\sin\frac{\theta_{i,k}}{2}\ket{0_i}
+
\cos\frac{\theta_{i,k}}{2}\ket{1_i},
\end{equation}
with
\begin{equation}
\mathcal{H}_{i,k}^{\mathrm{loc}}\ket{\tilde{1}_{i,k}}
=
+\Omega_{i,k}\ket{\tilde{1}_{i,k}},
\qquad
\mathcal{H}_{i,k}^{\mathrm{loc}}\ket{\tilde{0}_{i,k}}
=
-\Omega_{i,k}\ket{\tilde{0}_{i,k}}.
\end{equation}
Here $\ket{0_i}$ and $\ket{1_i}$ denote the bare local basis associated with $\sigma_i^z$, and defined by $\sigma_i^z\ket{0_i}=+\ket{0_i}$, $\sigma_i^z\ket{1_i}=-\ket{1_i}$.

The local dissipator is therefore constructed from jump operators defined in the instantaneous eigenbasis of the local Hamiltonian:
\begin{equation}
\tilde{\sigma}_{i,k}^{-}
=
\ket{\tilde{0}_{i,k}}\bra{\tilde{1}_{i,k}},
\qquad
\tilde{\sigma}_{i,k}^{+}
=
\ket{\tilde{1}_{i,k}}\bra{\tilde{0}_{i,k}}.
\end{equation}
Or, equivalently, defining $\tilde{\sigma}_{i,k}^{x}=\cos\theta_{i,k}\sigma_i^x -\sin\theta_{i,k}\sigma_i^z$ and $\tilde{\sigma}_{i,k}^{y}=\sigma_i^y$, one has $\tilde{\sigma}_{i,k}^{\pm} =\frac{1}{2}\left(\tilde{\sigma}_{i,k}^{x} \pm i\tilde{\sigma}_{i,k}^{y}\right)$.

To simplify the notation, we now omit the indices $i,k$ in the following single-spin equations. The rotated local master equation during a fixed input interval reads
\begin{equation}
\dot{\rho}
=
-i
\left[
\Omega\tilde{\sigma}^{z},
\rho
\right]
+
\Gamma_{\downarrow}
\mathcal{D}
[
\tilde{\sigma}^{-}
]\rho
+
\Gamma_{\uparrow}
\mathcal{D}
[
\tilde{\sigma}^{+}
]\rho .
\end{equation}
The rates satisfy detailed balance with respect to the rotated local splitting $\Gamma_{\uparrow}/\Gamma_{\downarrow}=e^{-2\beta\Omega}$, and, writing the state in the dressed Bloch basis $\rho=\frac{1}{2}\left(\mathbb{I}+\tilde{x}\tilde{\sigma}^{x} +\tilde{y}\tilde{\sigma}^{y}+\tilde{z}\tilde{\sigma}^{z}\right)$, the Bloch equations are
\begin{align}
\dot{\tilde{x}}
&=
-2\Omega\tilde{y}
-
\frac{\Gamma_{\downarrow}+\Gamma_{\uparrow}}{2}
\tilde{x},
\\
\dot{\tilde{y}}
&=
2\Omega\tilde{x}
-
\frac{\Gamma_{\downarrow}+\Gamma_{\uparrow}}{2}
\tilde{y},
\\
\dot{\tilde{z}}
&=
-
\left(
\Gamma_{\downarrow}
+
\Gamma_{\uparrow}
\right)
\tilde{z}
+
\Gamma_{\uparrow}
-
\Gamma_{\downarrow}.
\end{align}
Therefore, the stationary solution is
\begin{equation}
\tilde{x}_{\mathrm{ss}}=0,
\qquad
\tilde{y}_{\mathrm{ss}}=0,
\qquad
\tilde{z}_{\mathrm{ss}}
=
\frac{
\Gamma_{\uparrow}
-
\Gamma_{\downarrow}
}{
\Gamma_{\uparrow}
+
\Gamma_{\downarrow}
}
=
-\tanh(\beta\Omega).
\end{equation}
In the zero-temperature limit $\tilde{z}_{\mathrm{ss}}\rightarrow -1$, so the spin relaxes to the instantaneous ground state. Restoring the indices, the stationary polarizations in the original Pauli basis are
\begin{align}
\langle\sigma_{i}^x\rangle_{\mathrm{ss}}
&=
\tilde{z}_{\mathrm{ss}}\,\sin\theta_{i,k}
=
-
\frac{a_k}{\Omega_{i,k}}
\tanh\left(\beta\Omega_{i,k}\right),
\\
\langle\sigma_i^y\rangle_{\mathrm{ss}}
&=
0,
\\
\langle\sigma_i^z\rangle_{\mathrm{ss}}
&=
\tilde{z}_{\mathrm{ss}}\,
\cos\theta_{i,k}
=
-
\frac{h_i}{\Omega_{i,k}}
\tanh\left(\beta\Omega_{i,k}\right).
\end{align}
At zero temperature this reduces to
\begin{equation}
\langle\sigma_i^x\rangle_{\mathrm{ss}}
=
-\frac{a_k}{\sqrt{h_i^2+a_k^2}},
\qquad
\langle\sigma_i^y\rangle_{\mathrm{ss}}
=
0,
\qquad
\langle\sigma_i^z\rangle_{\mathrm{ss}}
=
-\frac{h_i}{\sqrt{h_i^2+a_k^2}} .
\end{equation}
This rotated-basis result differs qualitatively from the stationary state obtained with fixed bare jumps $\sigma_i^\pm$. In particular, for $h_i\rightarrow0$ the bare-jump model gives $\langle\sigma_i^x\rangle_{\mathrm{ss}}\rightarrow0$, whereas the rotated thermal model gives $\langle\sigma_i^x\rangle_{\mathrm{ss}}\rightarrow -1$ at zero temperature. This is because the rotated dissipator relaxes the spin towards the ground state of the instantaneous driven Hamiltonian, which is aligned along the $x$ direction when the drive dominates.

In the limit in which the bare local field dominates the input drive, $|h_i|\gg a_k$, one has
\begin{equation}
\Omega_{i,k}
=
\sqrt{h_i^2+a_k^2}
\simeq
|h_i|
\left[
1+
\frac{a_k^2}{2h_i^2}
+
O\left(\frac{a_k^4}{h_i^4}\right)
\right].
\end{equation}
For $h_i>0$ and at zero temperature, the rotated-basis stationary polarizations therefore reduce to
\begin{equation}
\langle\sigma_i^x\rangle_{\mathrm{ss}}
\simeq
-\frac{a_k}{h_i}
+
O\left(\frac{a_k^3}{h_i^3}\right),
\qquad
\langle\sigma_i^z\rangle_{\mathrm{ss}}
\simeq
-1
+
\frac{a_k^2}{2h_i^2}
+
O\left(\frac{a_k^4}{h_i^4}\right),
\end{equation}
while $\langle\sigma_i^y\rangle_{\mathrm{ss}}=0$. Thus, in the regime $h_i\gg a_k$, the rotated-basis thermal dissipator recovers the same leading behaviour obtained with the bare local jumps for the dominant $x$ and $z$ components: the transverse response scales as $-a_k/h_i$, while the spin remains almost polarized along the local $z$ direction. The two descriptions differ only at subleading order. In particular, with bare jumps, the dissipative basis remains slightly misaligned with the instantaneous Hamiltonian basis, producing a small quadrature component $\langle\sigma_i^y\rangle_{\rm ss}\simeq \gamma a_k/(4h_i^2)$. With rotated jumps, the steady state is diagonal in the instantaneous energy basis and the Bloch vector lies in the $xz$ plane, giving $\langle\sigma_i^y\rangle_{\rm ss}=0$. This difference vanishes in the large-$h_i$ limit and is therefore subleading.
\subsection{Work estimate with jump operators in the rotated local basis}
We now estimate the switching work when the local jump operators are defined in the instantaneous rotated basis of each single-spin Hamiltonian. For simplicity, we focus on the zero-temperature limit. The work associated with the instantaneous switch $a_k\rightarrow a_{k+1}$ is still
\begin{equation}
\mathcal{W}_k
=
(a_{k+1}-a_k)
\sum_{i=1}^N
\langle\sigma_i^x\rangle_{\rho_k}.
\end{equation}
Approximating $\rho_k$ by the product of the local stationary states in the rotated basis corresponding to the input $a_k$, and defining for simplicity 
\begin{equation}
F(a,h)
\equiv
\frac{a}{\sqrt{h^2+a^2}},
\end{equation}
one obtains
\begin{equation}
\mathcal{W}_k
\approx
-
(a_{k+1}-a_k)
\sum_{i=1}^N F(a_k,h_i).
\end{equation}
The total work over a sequence of $M$ inputs is therefore
\begin{equation}
\mathcal{W}_{\rm tot}
=
\sum_{k=1}^{M-1} \mathcal{W}_k
=
-
\sum_{k=1}^{M-1}
(a_{k+1}-a_k)
\sum_{i=1}^N F(a_k,h_i).
\end{equation}
Equivalently, if the input sequence is sampled independently from the same stationary distribution $p(a)$ at each step, all switching events are statistically equivalent. Therefore,
\begin{equation}
\mathbb{E}_a[\mathcal{W}_{\rm tot}]
=
\sum_{k=1}^{M-1}\mathbb{E}_a[\mathcal{W}_k]
=
(M-1)\mathbb{E}_a[\mathcal{W}_k].
\end{equation}
Thus, once the average work per switch has been computed, the average total work follows simply by multiplying by the number of switches. This is the reason why, in the following, we focus on $\mathbb{E}_a[\mathcal{W}_k]$.
\subsubsection{Homogeneous case, work per switch}
In the homogeneous case, $\delta_i=0$, so that $h_i=h$ for all spins. The work per switch becomes
\begin{equation}
\mathcal{W}_k
=
-
N(a_{k+1}-a_k) F(a_k,h).
\end{equation}
In the weak-input regime, $h\gg a_k$, one has $F(a_k,h)\simeq\frac{a_k}{h} \left[1-\frac{a_k^2}{2h^2}+O\left(\frac{a_k^4}{h^4}\right)\right]$, and therefore
\begin{equation}
\mathcal{W}_k
\simeq
-
\frac{N}{h}
(a_{k+1}-a_k)a_k
+
O\left(\frac{a_k^3(a_{k+1}-a_k)}{h^3}\right).
\end{equation}
Thus, for $h\gg a_k$, the rotated-basis estimate recovers the same leading large-$h$ scaling obtained with the bare local jumps. In the opposite strong-input regime, $h\ll a_k$, $F(a_k,h)\simeq1-\frac{h^2}{2a_k^2}+O\left(\frac{h^4}{a_k^4}\right)$. Hence,
\begin{equation}
\mathcal{W}_k
\simeq
-
N(a_{k+1}-a_k)
+
N(a_{k+1}-a_k)
\frac{h^2}{2a_k^2}
+
O\left(\frac{h^4(a_{k+1}-a_k)}{a_k^4}\right).
\end{equation}
In this limit the spins are almost fully polarized along the negative $x$ direction, so the transverse polarization no longer increases with the input amplitude.
\subsubsection{Disordered case, work per switch}
For uniformly distributed disorder, $\delta_i\in[-D,D]$, the disorder-averaged
single-spin response at fixed input amplitude is
\begin{equation}
\mathbb{E}_\delta[F(a,h+\delta)]
=
\frac{1}{2D}
\int_{-D}^{D}
d\delta\,
\frac{a}{\sqrt{(h+\delta)^2+a^2}}
=
\frac{a}{2D}
\left[
\sinh^{-1}\left(\frac{h+D}{a}\right)
-
\sinh^{-1}\left(\frac{h-D}{a}\right)
\right].
\end{equation}
Therefore,
\begin{equation}
\mathbb{E}_\delta[\mathcal{W}_k]
=
-
N(a_{k+1}-a_k)\mathbb{E}_\delta[F(a,h+\delta)]
=
-
N(a_{k+1}-a_k)
\frac{a_k}{2D}
\left[
\sinh^{-1}\left(\frac{h+D}{a_k}\right)
-
\sinh^{-1}\left(\frac{h-D}{a_k}\right)
\right].
\end{equation}
For weak disorder, $D\ll h$, one can expand around the homogeneous result:
\begin{equation}
\mathbb{E}_\delta[F(a,h+\delta)]
\simeq
F(a,h)
+
\frac{D^2}{6}
\frac{\partial^2 F(a,h)}{\partial h^2}
+
O(D^4)
=
\frac{a}{\sqrt{h^2+a^2}}
+
\frac{D^2}{6}
\frac{a(2h^2-a^2)}{(h^2+a^2)^{5/2}}
+
O(D^4).
\end{equation}
This expression shows that the effect of weak disorder depends on the ratio $h/a$. For $h\gg a$, the correction is positive and one recovers
\begin{equation}
\mathbb{E}_\delta[\mathcal{W}_k]
\simeq
-N(a_{k+1}-a_k)
\frac{a_k}{h}
\left[
1+\frac{D^2}{3h^2}
+
O\left(\frac{D^4}{h^4}\right)
+
O\left(\frac{a_k^2}{h^2}\right)
\right].
\end{equation}
Thus, in the weak-input regime, weak disorder slightly enhances the average transverse response because some spins have smaller effective detuning.

When $h=O(a)$, the correction remains of order $D^2/h^2$, but its sign and magnitude depend on the precise value of $h/a$. In particular, the leading weak-disorder correction changes sign at $h={a}/{\sqrt{2}}$. Therefore, around the crossover $h\sim a$, weak disorder can either increase or decrease the local transverse response, and its effect is not captured by the large-$h$ expansion alone.

In the strong-input regime, $h\ll a$, one finds instead
\begin{equation}
\mathbb{E}_\delta[\mathcal{W}_k]
\simeq
-N(a_{k+1}-a_k)
\left[
1
-
\frac{h^2}{2a_k^2}
-
\frac{D^2}{6a_k^2}
+
O\left(\frac{h^4}{a_k^4}\right)
+
O\left(\frac{h^2D^2}{a_k^4}\right)
+
O\left(\frac{D^4}{a_k^4}\right)
\right].
\end{equation}
Thus, when the input drive dominates the local field, weak disorder reduces the average transverse response: the spin is already almost aligned along the input direction, and fluctuations in the longitudinal field $h+\delta$ move it slightly away from perfect $x$-polarization. Hence, in the rotated-basis model, weak disorder has qualitatively different effects in the weak-input and strong-input regimes: it enhances the response for $h\gg a_k$, can change sign around $h\sim a_k$, and suppresses the response for $h\ll a_k$.

In the opposite strong-disorder regime, $D\gg h,a_k$, the response is suppressed because most spins have large $|h_i|$. From the exact disorder average one finds the asymptotic behaviour
\begin{equation}
\mathbb{E}_\delta[\mathcal{W}_k]
\simeq
-N(a_{k+1}-a_k)
\left[
\frac{a_k}{D}
\ln\left(\frac{2D}{a_k}\right)
-
\frac{a_k h^2}{2D^3}
+
O\left(\frac{a_k^3}{D^3}\right)
+
O\left(\frac{a_k h^4}{D^5}\right)
\right].
\end{equation}
Thus, the leading strong-disorder behaviour is independent of the average field $h$; the dependence on $h$ appears only in subleading corrections of order $h^2/D^2$ relative to the leading term. In contrast to the bare-jump model, there is no cancellation between positive and negative effective fields $h_i$, because the rotated-basis transverse response depends on $h_i^2$. Strong disorder suppresses the work because most spins are far detuned from the input scale.
\subsubsection{Input-averaged work}
We now compute the work averaged over independent uniformly distributed inputs. Let $a=A(s+1)$ with $s\in[0,1]$ uniformly distributed. Then $p(a)=1/A$ and $\mathbb{E}[a]=3A/2$. For fixed $h_i$, and since $a_{k+1}$ is independent of $a_k$,
\begin{equation}
\mathbb{E}_a[\mathcal{W}_k]
=
-\sum_{i=1}^N
\mathbb{E}_a
\left[
(a_{k+1}-a_k)F(a_k,h_i)
\right]
=
\sum_{i=1}^N
\mathbb{E}_a\left[aF(a,h_i)\right]
-
\mathbb{E}_a[a]\mathbb{E}_a\left[F(a,h_i)\right]
=
\sum_{i=1}^N
\mathrm{Cov}_a
\left[
a,
F(a, h_i)
\right].
\end{equation}
Since $F(a,h_i)$ is an increasing function of $a$, this average work is positive. For a single spin with local field $h_i$, we introduce
\begin{equation}
I_1(h_i;A)
=
\frac{1}{A}
\int_A^{2A}
da\,
\frac{a}{\sqrt{h_i^2+a^2}}
=
\frac{1}{A}
\left[
\sqrt{h_i^2+4A^2}
-
\sqrt{h_i^2+A^2}
\right],
\end{equation}
and
\begin{equation}
I_2(h_i;A)
=
\frac{1}{A}
\int_A^{2A}
da\,
\frac{a^2}{\sqrt{h_i^2+a^2}}
=
\frac{1}{2A}
\left[
2A\sqrt{h_i^2+4A^2}
-
A\sqrt{h_i^2+A^2}
-
h_i^2
\ln
\left(
\frac{
2A+\sqrt{h_i^2+4A^2}
}{
A+\sqrt{h_i^2+A^2}
}
\right)
\right].
\end{equation}
Therefore, the average work per switch for one spin is
\begin{equation}
\mathbb{E}_a[\mathcal{W}_{k,i}]
=
I_2(h_i;A)
-
\frac{3A}{2}I_1(h_i;A).
\end{equation}
\subsubsection{Homogeneous input-averaged work}
In the homogeneous case, $h_i=h$, the average work per switch is
\begin{equation}
\mathbb{E}_a[\mathcal{W}_k]
=
N
\left[
I_2(h;A)
-
\frac{3A}{2}I_1(h;A)
\right].
\end{equation}
For $h\gg A$, one has $F(a,h)\simeq a/h$. Therefore, the average work per
switch becomes
\begin{equation}
\mathbb{E}_a[\mathcal{W}_k]
=
N\,\mathrm{Cov}_a\left[a,F(a,h)\right]
\simeq
\frac{N}{h}\mathrm{Cov}_a\left[a,a\right]
=
\frac{N}{h}\mathrm{Var}(a)
=
\frac{N A^2}{12h}.
\end{equation}
Thus, in the weak-input regime, the average work scales as $A^2/h$, as in the bare-jump estimate. For $h\ll A$, one has $F(a,h)\simeq 1-\frac{h^2}{2a^2} +O\left(\frac{h^4}{a^4}\right)$. Thus, since $\mathrm{Cov}_a\left[a,1\right]=0$, the first nonzero contribution to the input-averaged work is then
\begin{equation}
\mathrm{Cov}_a\left[a,F(a,h)\right]
\simeq
-\frac{h^2}{2}
\mathrm{Cov}_a\left[a,\frac{1}{a^2}\right]
=
-\frac{h^2}{2}
\left(
\mathbb{E}_a\left[\frac{1}{a}\right]
-
\mathbb{E}_a[a]\,
\mathbb{E}_a\left[\frac{1}{a^2}\right]
\right)
=
-\frac{h^2}{2A}
\left(
\ln 2-\frac{3}{4}
\right).
\end{equation}
Since $\ln2<3/4$, this work is positive. Therefore,
\begin{equation}
\mathbb{E}_a[\mathcal{W}_k]
\simeq
\frac{N h^2}{2A}
\left(
\frac{3}{4}
-
\ln 2
\right)
+
O\left(\frac{h^4}{A^3}\right).
\end{equation}
Thus, in the strong-input regime $h\ll A$, the average work vanishes as $h^2/A$. Physically, the spin is almost fully polarized along the negative $x$ direction, so the transverse response is nearly independent of the input amplitude.

\subsubsection{Disorder-averaged input-averaged work}
For a fixed disorder realization, the input-averaged work is
\begin{equation}
\mathbb{E}_a[\mathcal{W}_k]
=
\sum_{i=1}^N
\left[
I_2(h_i;A)
-
\frac{3A}{2}I_1(h_i;A)
\right],
\qquad
h_i=h+\delta_i.
\end{equation}
Averaging also over the disorder gives
\begin{equation}
\mathbb{E}_{\delta,a}[\mathcal{W}_k]
=
N\,\mathbb{E}_D\left[G(h+\delta;A)\right],
\end{equation}
where we have defined
\begin{equation}
G(h;A)
\equiv
I_2(h;A)
-
\frac{3A}{2}I_1(h;A)
=
\frac{1}{A}
\int_A^{2A} da\,
\frac{a^2-\frac{3A}{2}a}
{\sqrt{h^2+a^2}}.
\end{equation}
Therefore,
\begin{equation}
\mathbb{E}_{\delta,a}[\mathcal{W}_k]
=
\frac{N}{2D}
\int_{-D}^{D}
d\delta\,
G(h+\delta;A).
\end{equation}
By exchanging the disorder and input integrals, this can also be written as
\begin{equation}
\mathbb{E}_{\delta,a}[\mathcal{W}_k]
=
\frac{N}{2DA}
\int_A^{2A} da\,
\left(
a^2-\frac{3A}{2}a
\right)
\left[
\sinh^{-1}\left(\frac{h+D}{a}\right)
-
\sinh^{-1}\left(\frac{h-D}{a}\right)
\right].
\label{eq:Sup_main_switch_work}
\end{equation}
Since the input and disorder averages commute, we can start directly from the asymptotic expressions for $\mathbb{E}_\delta[\mathcal{W}_k]$ and average them over independent inputs $a_k,a_{k+1}\in[A,2A]$.

In the weak-disorder and weak-input regime, $D\ll h$ and $h\gg a_k$, we have
\begin{equation}
\mathbb{E}_{\delta,a}[\mathcal{W}_k]
\simeq
\frac{N A^2}{12h}
\left[
1+\frac{D^2}{3h^2}
+
O\left(\frac{D^4}{h^4}\right)
+
O\left(\frac{A^2}{h^2}\right)
\right].
\end{equation}
For the weak-disorder and strong-input regime,
since $\mathbb{E}_a[a_{k+1}-a_k]=0$, the constant term gives no contribution after averaging. The first nonzero contribution is therefore
\begin{equation}
\mathbb{E}_{\delta,a}[\mathcal{W}_k]
\simeq
N\left(
\frac{h^2}{2}+\frac{D^2}{6}
\right)
\mathbb{E}_a\left[
\frac{a_{k+1}-a_k}{a_k^2}
\right]
=
N\left(
\frac{h^2}{2}+\frac{D^2}{6}
\right)
\left(
\mathbb{E}_a[a]\,
\mathbb{E}_a\left[\frac{1}{a^2}\right]
-
\mathbb{E}_a\left[\frac{1}{a}\right]
\right).
\end{equation}
For $a\sim U[A,2A]$, $\mathbb{E}_a[a]=3A/2$, $\mathbb{E}_a\left[1/a^2\right]=1/(2A^2)$ and $\mathbb{E}_a\left[1/a\right]=\ln2/A$. Hence,
\begin{equation}
\mathbb{E}_{D,a}[\mathcal{W}_k]
\simeq
\frac{N}{2A}
\left(
h^2+\frac{D^2}{3}
\right)
\left(
\frac{3}{4}-\ln2
\right).
\end{equation}
Thus, in the strong-input regime, the average work vanishes as $(h^2+D^2/3)/A$. The term proportional to $D^2$ appears because disorder fluctuations in the longitudinal field slightly reduce the almost perfect alignment along the input direction and restore a weak input dependence of the transverse response.

Finally, in the strong-disorder regime, $D\gg h,a_k$, averaging the leading term over the input distribution gives
\begin{equation}
\mathbb{E}_{\delta,a}[\mathcal{W}_k]
\simeq
-\frac{N}{D}
\mathbb{E}_a
\left[
(a_{k+1}-a_k)a_k
\ln\left(\frac{2D}{a_k}\right)
\right].
\end{equation}
Using independence,
\begin{equation}
\mathbb{E}_a
\left[
(a_{k+1}-a_k)a_k
\ln\left(\frac{2D}{a_k}\right)
\right]
=
\mathbb{E}_a[a]\,
\mathbb{E}_a\left[
a\ln\left(\frac{2D}{a}\right)
\right]
\nonumber
-
\mathbb{E}_a\left[
a^2\ln\left(\frac{2D}{a}\right)
\right].
\end{equation}
For $a\sim U[A,2A]$, this gives
\begin{equation}
\mathbb{E}_{\delta,a}[\mathcal{W}_k]
\simeq
\frac{N A^2}{12D}
\ln\left(\frac{D}{A}\right),
\end{equation}
up to constants inside the logarithm and subleading corrections of order $h^2/D^3$ and $A^3/D^3$. Thus, strong disorder suppresses the input-averaged work approximately as $(\ln D)/D$. This suppression is not due to sign cancellations between positive and negative effective fields, but to the fact that most spins have large $|h_i|$ and therefore weak transverse response.

The estimates above assume that the system has enough time to approach the local rotated-basis stationary state during each input interval, and they neglect interaction-induced corrections. However, as discussed in the main text, interactions may enhance local relaxation and improve the quality of the stationary approximation.
For the actual simulations with discretized inputs, the integrals over $a\in[A,2A]$ should be replaced by finite sums over the input grid. The continuous formulas are useful analytical benchmarks and are recovered in the large-grid limit.
\section*{Additional interaction strengths}
In this section we report the disorder and input-amplitude scans for additional values of the interaction strength, complementing the representative results presented in the End Matter of the main text.

\begin{figure}[t!]
    \includegraphics[width=\textwidth]{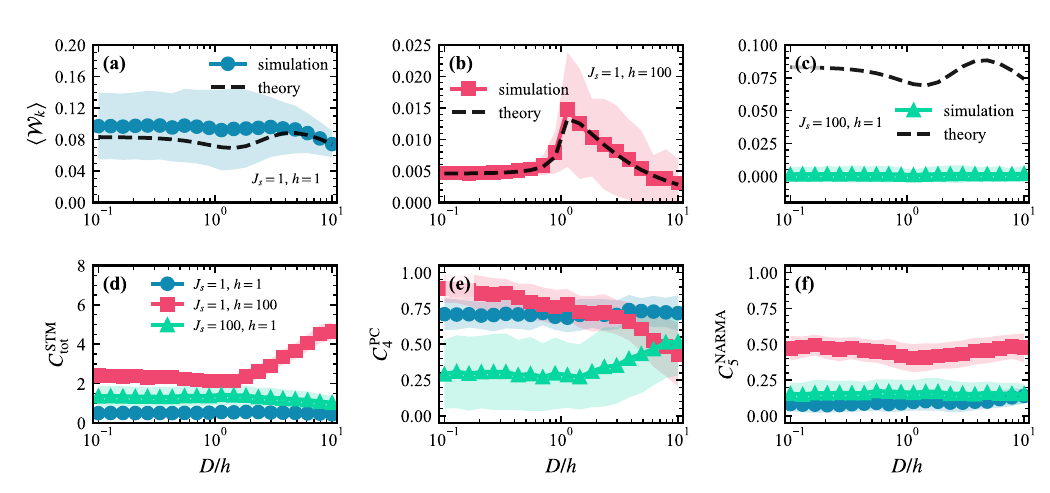}
	\caption{Dependence on the disorder strength $D/h$ for three representative interaction strengths, $J_s/h=10^{-2}$, $1$, and $10^{2}$. Panels (a--c) show the average switching work together with the analytical prediction of Eq.~\eqref{eq:Sup_main_switch_work}. Panels (d--f) report the average capacities of the STM, fourth-order PC, and fifth-order NARMA tasks, respectively.}
    \label{fig:S1}
\end{figure}

\begin{figure}[t!]
    \includegraphics[width=\textwidth]{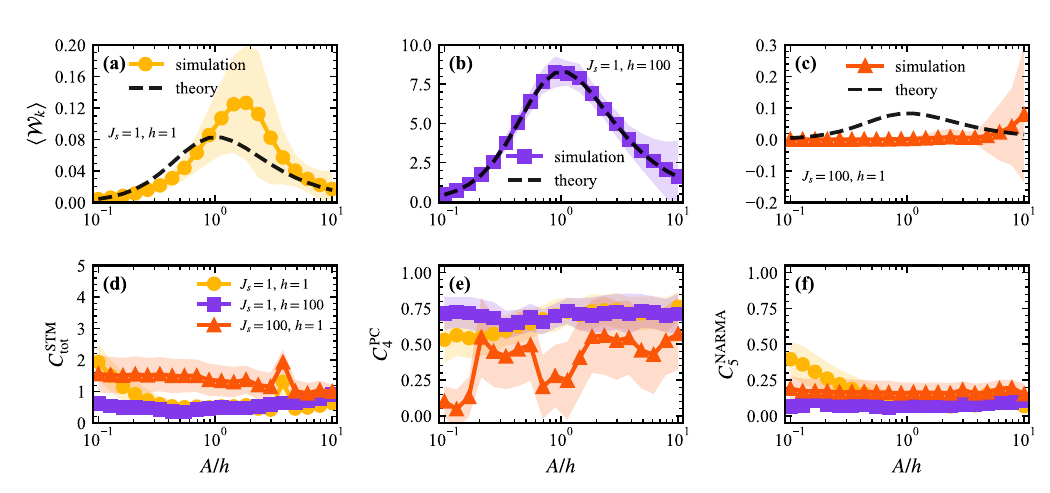}
	\caption{Dependence on the input amplitude $A/h$ for three representative interaction strengths, $J_s/h=10^{-2}$, $1$, and $10^{2}$. Panels (a--c) show the average switching work together with the analytical prediction of Eq.~\eqref{eq:Sup_main_switch_work}. Panels (d--f) report the average capacities of the STM, fourth-order PC, and fifth-order NARMA tasks, respectively.}
    \label{fig:S2}
\end{figure}

Figures~\ref{fig:S1} and \ref{fig:S2} show the disorder and input-amplitude scans for three representative interaction strengths, $J_s/h=10^{-2}$, $1$, and $10^{2}$. As expected, the analytical prediction for the switching work remains accurate only in the perturbative regime $J_s/h\ll1$, where the assumptions underlying Eq. \ref{eq:Sup_main_switch_work} are satisfied. A modest enhancement of the fourth-order parity-check capacity is observed at large $J_s/h$ upon increasing either the disorder strength or the input amplitude.


\end{document}